\documentclass{article}

\usepackage{arxiv}

\usepackage[utf8]{inputenc} 
\usepackage[T1]{fontenc}    
\usepackage{hyperref}       
\usepackage{url}            
\usepackage{booktabs}       
\usepackage{amsfonts}       
\usepackage{nicefrac}       
\usepackage{microtype}      
\usepackage{lipsum}		
\usepackage{graphicx}
\usepackage{natbib}
\usepackage{doi}

\usepackage{amsmath}
\usepackage{algorithm}
\usepackage{algpseudocode}

\title{Position bias in features}

\author{ {\hspace{1mm}Richard Demsyn-Jones} \\
	Thumbtack \\
	\texttt{demsynjones@thumbtack.com} \\
	\texttt{rdemsynjones@gmail.com}
}

\renewcommand{\headeright}{}
\renewcommand{\undertitle}{}

\hypersetup{
pdftitle={Position bias in features},
pdfauthor={Richard Demsyn-Jones},
pdfkeywords={position bias, search ranking, learning to rank, inverse propensity scoring, propensity estimation},
}

\begin{document}
\maketitle

\begin{abstract}
The purpose of modeling document relevance for search engines is to rank better in subsequent searches. Document-specific historical click-through rates can be important features in a dynamic ranking system which updates as we accumulate more sample. This paper describes the properties of several such features, and tests them in controlled experiments. Extending the inverse propensity weighting method to documents creates an unbiased estimate of document relevance. This feature can approximate relevance accurately, leading to near-optimal ranking in ideal circumstances. However, it has high variance that is increasing with respect to the degree of position bias. Furthermore, inaccurate position bias estimation leads to poor performance. Under several scenarios this feature can perform worse than biased click-through rates. This paper underscores the need for accurate position bias estimation, and is unique in suggesting simultaneous use of biased and unbiased position bias features.
\end{abstract}


\section{Introduction}

This paper takes a fundamentally predictive and feature-centric view of learning-to-rank. In applied settings, the purpose of modeling relevance is to rank better in subsequent searches. Document-specific historical click-through rates can be important features in a dynamic ranking system which updates as we accumulate more sample.

The main contributions of this work are:

\begin{itemize}
    \item Splitting from the literature, which emphasizes \textit{query-\\document} relevance, to establish the notion and usefulness of \textit{document} relevance.
    \item Formally defining and establishing the properties of five click-through features: click through rate (CTR), inverse propensity weighted click-through rate (IPW-CTR), click-through rate weighted by observed click rates by position (Empirical-CTR), clicks over expected clicks (COEC, or \\Empirical-COEC), and inverse propensity weighted clicks over expected clicks (IPW-COEC). In particular, the COEC features are rarely covered in prior literature.
    \item Extending a small literature that shows how biased low variance estimators can create better performance for learning-to-rank than unbiased methods.
    \item Applying controlled experiments to show the sensitivity of different features to sample size and degree of position bias. Notably, empirical weights lead to very poor ranking, such that accurate estimation of position bias is more important than the choice between IPW, COEC, or SNIPS weighting schemes.
\end{itemize}

The motivation for this work is an empirical setting where historical click-through rates—of various forms—have continually been used as features in ranking models. These features have persisted over many iterations of production rankers which used multiple types of models. Despite the availability of many direct features of ranked documents, click-through rate features have been among the strongest features by a variety of typical feature importance metrics, and improvements to these features have improved model performance metrics and been validated on downstream business metrics in user-randomized experiments.

\subsection{Notation}

See Table \ref{table:symbols} for a description of symbols used throughout this text.

\begin{table}
\centering
  \caption{Symbols used in this text}
\begin{tabular}{cll}
\toprule
    Symbol&Meaning\\
    \midrule
$R$        & Relevance                                  \\
$E$        & Examination                                \\
$I$        & The indicator function                      \\
$\mathcal{L}$       & Global dataset                             \\
$N$        & Size of global dataset                     \\
$\theta$    & Unique sequence of position bias factors   \\
$Q$        & Unique sequence of possible queries        \\
$C$        & Click, Unique sequence of possible click outcomes \\
$D$        & Unique sequence of possible documents      \\
$d$        & Document \\
$l$        & Document dataset                           \\
$n$        & Size of document dataset                   \\
$q$        & Query, Vector of queries for document             \\
$c$        & Click, Vector of clicks for document              \\
$p$        & Position, Vector of positions for document           \\
$\vartheta$ & Vector of position biases for document     \\
$\delta$    & Any specific document                      \\
$c^\prime$  & Vector of clicks across all documents      \\
$p^\prime$  & Vector of positions across all documents  \\
\bottomrule
\end{tabular}
\label{table:symbols}
\end{table}

For convenience, the notation $P(X)$ for any random binary variable $X$ is shorthand for $P(X=1)$. When needed, the probability of the random variable equaling zero will be explicitly written as $P(X=0)$.

\section{Previous work}

Search engines usually return a multitude of results, presenting them in an ordered list. The act of determining the order, known as \textit{ranking}, is an essential challenge. Users do not select items randomly from the result list, but rather have a strong preference for the top-most (or, sometimes, left-most) positions \citep{Joachims07}. This \textit{position bias} is a central problem to modeling optimal rank, known as the \textit{learning-to-rank} problem, and is well-studied \citep{Chapelle09, Craswell08, Dupret08, Joachims17, Ai2018, Agarwal2018, Wang18, Aslanyan19, Agarwal19, Agarwal19b, Saito2019, Vardasbi20, Ovaisi20, Yan22, Zhang22}.

The position-based propensity model (PBM) is a common behavior model used in 
learning-to-rank. In the PBM, positive user choices (known in the literature as \textit{click}, $C$, which could represent any desired user action) depend on the probability that the user examines ($E$) the document and the probability that an item is relevant ($R$) to the user's query. These two probabilities are multiplicative, as shown in Equations \ref{eq:pbm_assumption} and \ref{eq:pbm_base}, and the probability of examination depends only on position ($k$). This is the \textit{separability hypothesis} \citep{Dupret08}.

\begin{equation}
  C = 1 \Leftrightarrow E = 1, R = 1
  \label{eq:pbm_assumption}
\end{equation}

\begin{equation}
  P(C | query, document, position) = P(R|query, document) \cdot P(E|position)
  \label{eq:pbm_base}
\end{equation}

The literature defines $\theta_k = P(E|k)$, such that $\theta$ is a vector of position biases for different positions.

Machine learning algorithms can be unbiased estimators of the PBM if fit with inverse propensity scoring (IPS), also known as inverse propensity weighting (IPW). We will use "IPW" in this text for natural English interpretation of our feature names. In IPW each positive observation is weighted by the inverse likelihood of examination, such that positive observations that were at low positions in the corpus are given higher weight than those from higher positions.

Other methods directly optimize all parameters of the relevance model, estimating position bias factors simultaneously with identifying relevance factors. Typically these models can be fit using an Expectation-Maximization framework \citep{Wang18, Agarwal19b}. Alternatively, these jointly-optimized models are commonly fit in neural network architectures \citep{Agarwal2018, Ai2018, Guo19, Yan22, Zhang22}.

Our paper will be agnostic about the model architecture used, presuming that practitioners choose an architecture fitting their setting and beliefs. The insights of this paper about position bias in features are relevant to features that could be used across architectures.

This paper uses a basic setting with a total ranking (e.g. a never-ending pagination of results) and asymptotic behavior where even distant ranked results receive some clicks. This can be contrasted with settings where lists are truncated early or search volume is insufficient relative to the number of relevant documents. If there is a sufficiently large number of documents with zero clicks, then this \textit{cold start} or \textit{long tail} problem can be a first-order concern. This paper fits into a large literature that emphasizes the challenge of position bias correction even when all documents have some exposure to users and have corresponding click data, when cold start is a secondary concern relative to dramatic position bias within the early part of search results. Furthermore this paper discusses and demonstrates ranking with heterogeneous exposure across documents.

\subsection{Document relevance}

We choose to use a concept of document relevance, defined for a document $\delta$ as $E\left[R|D=\delta\right]$. This is not a typical concept in the literature, which instead defines \textit{query-document} relevance, such that relevance is always in the eye of the beholder. Notably, two of the most popular datasets in the learning to rank literature contain no document identifiers and have no discussion of document recurrence across queries \citep{Chapelle10, Qin2010}.

Yet in our datasets outside of research domains—and anecdotally—we notice considerable correlation in relevance labels across users for the same document. Consider the extreme case of any item of obvious spam, which might have uniformly low query-document relevance. Our paper is agnostic about whether documents have intrinsic relevance with causal effects on query-document relevance or whether document relevance is simply an observational summary of query-document relevance across all queries. Regardless, this concept of document relevance has predictive power useful for subsequent ranking.

Much of the learning-to-rank literature focuses on unbiased ranking algorithms or consistent estimates of the effect of position bias itself. Our paper fits in a smaller literature emphasizing features for subsequent ranking derived from historical click data from searches with each document. Once we accept document relevance as a predictive concept, estimation of a document's relevance from prior data can be useful for subsequent ranking. These features "are good estimators of the future performance" \citep{Cheng2010}. 

One notable such feature is clicks over expected clicks (COEC), introduced in \cite{Zhang2007} as a successor to click difference from \cite{Agichtein2006}, and used later in \cite{Schroedl2010} and \cite{Liu2017}. COEC attempts to debias historical clicks by adjusting for click performance on all documents at each position. However at least one study found that COEC only "gave similar results" to using a biased click metric \citep{Teo2016}. Although all those papers define COEC, ours is the first known paper to demonstrate the properties of COEC formally. We will contrast COEC against other historical click features, explaining when it is more or less useful, and demonstrating these relationships with simulations.

\section{Biased and unbiased click-through rate (CTR) features}

Consider a search product where users conduct searches, such that for each search they submit a query to the search engine and the search engine returns an ordered list of documents. Documents may repeat across different searches. For simplicity and without loss of generality we will restrict to the case where documents are unique within a search.\footnote{While this is common in search engines, there may be some search engines that choose to show the same document multiple times.} We will call each individual query-document pairing a \textit{record}, such that there may be multiple records returned for a query.




$Q$, $D$, $C$, and $\theta$ are finite unique sequences representing all possible values for queries, documents, clicks, and position biases. Order is irrelevant for $Q$ and $D$, but $\theta$ will be ordered by position such that $\theta_k$ is the position bias for the $k^{th}$ position. This matches the common use of $\theta$ in the literature, as described earlier. Our relevance metric, clicks, is binary without loss of generality, such that $C = (0, 1)$.

We will define many equations referring to a corpus of records for any specific document. If a document $\delta$ is present in $n$ records, then let $q$ be a vector of queries for which $\delta$ was shown, $p$ be the vector of positions that $\delta$ was shown in for those queries, $\vartheta$ be the vector of position biases ($\vartheta_i = \theta_{p_i}$), and $c$ be the vector of click outcomes. The vectors all have the same length and their ordering is consistent. $z$ contains tuples with items from those vectors, and $l$ is the logged dataset of records. 

\begin{equation}
  l = \{z_i\}_i^n = \{(q_i, p_i, \vartheta_i, c_i)\}_i^n
\end{equation}

At times we will refer to a corresponding dataset $\mathcal{L}$ that contains all records across all documents ($l \in \mathcal{L}$), with length $N$ ($N \ge n$, and typically $N >> n$).

\subsection{Click-through rate}

Click-through rate (CTR) is a biased estimate of relevance. Consider the CTR for a document:

\begin{equation}
  \text{CTR} = \frac{1}{n}\sum_{i=1}^{n} c_i
  \label{eq:raw_coec}
\end{equation}

As is a typical result, expectation of the average equals the expectation of the random variable $c$:

\begin{equation}
\begin{aligned}
\mathbb{E}\left[\frac{1}{n} \sum_{i=1}^n c_i\right] & = \frac{1}{n}\mathbb{E}\left[\sum_{i=1}^n c_i\right] \\ 
& = \frac{1}{n} \sum_{i=1}^n \mathbb{E}[c_i] \\ 
& = \frac{1}{n} n \mathbb{E}[c_i] \\ 
& = \mathbb{E}[c_i] \\ 
& = \mathbb{E}[c] \\
\end{aligned}
\end{equation}

Given that the probability of click is the probability of relevance multiplied by the probability of examination, when the examination probabilities do not equal 1 then click rate is a biased estimator of relevance. More specifically, since all positions have at most 100\% examination probability ($\forall k \in \left[1, |\theta|\right]: \theta_k \le 1$), and if we assume at least some position bias exists such that some position has below 100\% examination probability ($\exists{k} \in \left[1, |\theta|\right]: \theta_k < 1$), then the probability of click is lower than the probability of relevance ($\exists{k} \in \left[1, |\theta|\right] : P(R|q,d) \cdot \theta_k < P(R|q,d)$) and average CTR will be (in expectation) lower than average relevance.

\subsection{Inverse propensity weighted CTR}

As well-established since \citet{Joachims17}, the expected value of inverse propensity weighted click rate across the dataset $\mathcal{L}$ equals the expected value of relevance. This reasoning extends to metrics over single documents.

For any fixed document, $\delta$, the expected value of the an inverse propensity weighted click-through rate is the expected value of that document's relevance.

\begin{equation}
\begin{aligned}
\mathbb{E}\left[\frac{1}{n} \sum_{i=1}^n \frac{c_i}{\vartheta_{p_i}}\right] & = \mathbb{E}\left[\frac{c}{\vartheta}\right] \\ 
& = \sum_{k=1}^{|\theta|}\sum_{l=1}^{|C|}\frac{C_l}{\theta_k}P(c=C_l,\vartheta=\theta_k|d=\delta) \\
& = \sum_{i=1}^{|Q|}\sum_{k=1}^{|\theta|}\sum_{l=1}^{|C|}\frac{C_l}{\theta_k}P(c=C_l|Q_i,d=\delta,\theta_k)P(Q_i, d=\delta, \theta_k) \\
& = \sum_{i=1}^{|Q|}\sum_{k=1}^{|\theta|}\frac{1}{\theta_k}P(C|Q_i,d=\delta,\theta_k)P(Q_i, d=\delta, \theta_k) && (c \in \{0,1\})\\
& = \sum_{i=1}^{|Q|}\sum_{k=1}^{|\theta|}\frac{1}{\theta_k}P(R|Q_i,d=\delta)\theta_kP(Q_i, d=\delta, \theta_k) \\
& = \sum_{i=1}^{|Q|}\sum_{k=1}^{|\theta|}P(R|Q_i,d=\delta)P(Q_i, D=\delta, \theta_k) \\
& = \sum_{i=1}^{|Q|}P(R|Q_i,d=\delta)P(Q_i, d=\delta) \\
& = \mathbb{E}[R|d=\delta]
\end{aligned}
\end{equation}

We will refer to this as IPW-CTR, in contrast to the biased and unadjusted click-through rate (CTR).

\subsection{IPW with empirical weights}

Some practitioners could choose to use click rates by position as an estimate of position bias. These are sometimes referred to as \textit{empirical} click rates \citep{Wang18}. These click rates are observable and calculable with simple averaging. As a measure of position bias, they are biased. Typically the search engine's ranking algorithm has some ability to rank more relevant documents higher. The observed click rate at any position is lower than relevance, and in particular a position's empirical click rate relative to the first position is more dramatic than the true position bias relative to the first position, because examination probabilities lower with position and so will relevance. Thus this value overestimates the scale of position bias in terms of the relative ratio of examination probabilities between positions. This has been demonstrated with randomization studies for several large consumer products \citep{Wang18, Agarwal19, DemsynJones22}.

$\theta$ is the vector of true position bias weights. Consider an alternative, incorrect vector of position weights $\hat{\theta}$ that overestimates the extent of position bias and is scaled such that $\hat{\theta}_1 = \theta_1$, $\forall k \in \theta: \hat{\theta}_k \le \theta_k$, and $\exists{k} : \hat{\theta}_k < \theta_k$. Assume also that every position has non-zero probability of occurrence, or else would be omitted from the $k$ range. Define Empirical-CTR as $\frac{1}{n} \sum_{i=1}^n \frac{c_i}{\hat{\theta}_{p_i}}$.

\begin{equation}
\begin{aligned}
\mathbb{E}\left[\frac{1}{n} \sum_{i=1}^n \frac{c_i}{\hat{\theta}_{p_i}}\right] & = \mathbb{E}\left[\frac{c}{\hat{\theta}}\right] \\ 
& = \sum_{i=1}^{|Q|}\sum_{k=1}^{|\theta|}\frac{1}{\hat{\theta}_k}P(R|Q_i,d=\delta)\theta_kP(Q_i, d=\delta, \theta_k) \\
& = \sum_{i=1}^{|Q|}\sum_{k=1}^{|\theta|}\frac{\theta_k}{\hat{\theta}_k}P(R|Q_i,d=\delta)P(Q_i, d=\delta, \theta_k) \\
\end{aligned}
\end{equation}

Given that $\forall k \in \theta: \frac{\theta_k}{\hat{\theta}_k} \ge 1$, and $\exists{k} : \frac{\theta_k}{\hat{\theta}_k} > 1$, and that every position has non-zero occurrence probability, this equation is strictly larger than the IPW-CTR expectation.

\begin{equation}
\mathbb{E}\left[\frac{1}{n} \sum_{i=1}^n \frac{c_i}{\hat{\theta}_{p_i}}\right] > \mathbb{E}\left[\frac{1}{n} \sum_{i=1}^n \frac{c_i}{\theta_{p_i}}\right]
\end{equation}

\begin{equation}
\mathbb{E}[\text{Empirical-CTR}] > \mathbb{E}[\text{IPW-CTR}]
\end{equation}

Since the right hand side equals the expected relevance of the document, the left hand side is biased upwards relative to relevance.

Note that not all documents are affected equally. Given that the ratio $\frac{\theta_k}{\hat{\theta}_k}$ need not be constant, the distribution of positions will affect the scale of biasedness, such that there is no constant scaling factor to convert Empirical-CTR to IPW-CTR. 

\subsection{Self-normalized risk estimators}

The Self-Normalized Inverse Propensity Scoring (SNIPS) estimator uses the sum of position bias weights as a "control variate" \citep{Swaminathan2015, Schnabel2016}, as shown below:

\begin{equation}
\begin{aligned}
  \text{SNIPS} 
  & = \frac{\sum_{i=1}^{n} \frac{c_i}{\vartheta_{p_i}}}{\sum_{i=1}^{n} \frac{1}{\vartheta_{p_i}}} \\ 
  & = \frac{\frac{1}{n}\sum_{i=1}^{n} \frac{c_i}{\vartheta_{p_i}}}{\frac{1}{n}\sum_{i=1}^{n} \frac{1}{\vartheta_{p_i}}} \\
  & = \frac{\text{IPW-CTR}}{\frac{1}{n}\sum_{i=1}^{n} \frac{1}{\vartheta_{p_i}}} \\
\end{aligned}
\end{equation}

The SNIPS estimator is inherently biased downwards in any situation where position bias is present, because it equals an unbiased estimator (IPW-CTR) divided by an average that is strictly positive and above 1.

\subsection{Clicks over expected clicks}

In the click over expected clicks (COEC) approach we sum clicks and divide by \textit{expected clicks}, which are how many clicks we would expect based on observed click rates at each position \citep{Zhang2007}.

Let $c^\prime$ and $p^\prime$  be click and position sequences for the entire dataset $\mathcal{L}$ rather than the document-specific dataset $l$. Using $I$ as the indicator function, define $\hat{\theta}_k$ as:

\begin{equation}
  \hat{\theta}_k = \frac{\sum_{i=1}^{N} I(p^\prime_i = k) \cdot c^\prime_i}{\sum_{i=1}^{N}I(p^\prime_i = k)}
  \label{eq:coec_rates}
\end{equation}

COEC for a document is:

\begin{equation}
  \text{COEC} = \frac{\sum_{i=1}^{n} c_i}{\sum_{i=1}^{n} \hat{\theta}_{p_i}}
  \label{eq:coec}
\end{equation}

This differs from Empirical-CTR because we divide by the sum of position click-through rates, rather than dividing each observation by its position click-through rate.

\subsection{Inverse propensity weighted COEC}

Consider an inverse propensity weighted COEC (IPW-COEC), using true $\theta$ in its denominator.

\begin{equation}
\begin{aligned}
\text{IPW-COEC} = \frac{\sum_{i=1}^{n} c_i}{\sum_{i=1}^{n} \theta_{p_i}}
\end{aligned}
\end{equation}

For a constant position distribution for a document, IPW-COEC is also an unbiased estimator of relevance.

Define $\chi$ as a $|\theta|$-length vector of number of clicks at each position, such that $\sum_{k=1}^{|\theta|} \chi_k = \sum_{i=1}^{n} c_i$ and $\chi_k = \sum_{i=1}^{n} c_i \cdot I(p_i = k)$. Consider $p$ to be a fixed (non-random) vector.

\begin{equation}
\begin{aligned}
\text{IPW-COEC} & = \frac{\sum_{i=1}^{n} c_i}{\sum_{i=1}^{n} \theta_{p_i}} \\
& = \frac{1}{\sum_{i=1}^{n} \theta_{p_i}} \sum_{k=1}^{|\theta|} n_k \cdot \bar{\chi} \\
\end{aligned}
\end{equation}

\begin{equation}
\begin{aligned}
\mathbb{E}\left[\text{IPW-COEC}\right] & = \frac{1}{\sum_{i=1}^{n} \theta_{p_i}} \mathbb{E}\left[\sum_{i=1}^{|\theta|} \chi_k\right] \\
& = \frac{1}{\sum_{i=1}^{n} \theta_{p_i}} \sum_{i=1}^{|\theta|}\mathbb{E}\left[\chi_k\right] \\ 
& = \frac{1}{\sum_{i=1}^{n} \theta_{p_i}} \sum_{i=1}^{|\theta|} n_k P(R|d=\delta)\theta_k \\
& = P(R|d=\delta) \frac{\sum_{i=1}^{|\theta|} n_k \theta_k}{\sum_{i=1}^{n} \theta_{p_i}} \\
& = P(R|d=\delta) \frac{\sum_{i=1}^{n} \theta_{p_i}}{\sum_{i=1}^{n} \theta_{p_i}} \\
& = P(R|d=\delta)
\end{aligned}
\end{equation}

In general, IPW-COEC does not equal IPW-CTR. However the two features are equivalent when position is fixed, when there is some $k$ such that $\forall i \in n: p_i = k$. 

\begin{equation}
\begin{aligned}
\text{IPW-COEC} & = \frac{\sum_{i=1}^{n} c_i}{\sum_{i=1}^{n} \theta_{p_i}} \\
& = \frac{\sum_{i=1}^{n} c_i}{n \cdot \theta_k} && (\text{when position is fixed at $k$}) \\
& = \frac{1}{n} \cdot \sum_{i=1}^{n} \frac{c_i}{\theta_k} \\
& = \frac{1}{n} \cdot \sum_{i=1}^{n} \frac{c_i}{\theta_i} \\
& = \text{IPW-CTR}
\end{aligned}
\end{equation}

\section{The cost of unbiasedness}

Click-through rates are biased, but they have low variance. With clicks being binary outcomes, the vector of clicks for a document has a variance of $P(c) \cdot (1-P(c))$, which can be no larger than $\frac{1}{4}$. Given independent searches and Bienaymé's identity, the mean has a variance of $\frac{Var(c)}{n}$, which can be no larger than $\frac{1}{4n}$. 

By definition, $\forall k \in [1,|\theta|]: 0 < \theta_k \le 1$. IPW-CTR has a lower bound at 0, but has no upper bound: $\lim_{\theta_k\to0} \frac{1}{\theta_k} = \infty$. We can have very large ratios when we have very low $\theta_k$. Since IPW-CTR is a sum over these ratios, a single low-likelihood click can lead to an extreme IPW-CTR for any given document, even with the averaging.

\begin{equation}
\begin{aligned}
Var\left(\frac{c}{\theta}\right) & = \mathbb{E}\left[\left(\frac{c}{\theta}\right)^2\right] - \mathbb{E}\left[\frac{c}{\theta}\right]^2 \\
& = \mathbb{E}\left[\frac{c}{\theta^2}\right] - \mathbb{E}\left[R|D=\delta\right]^2 && (c^2 = c \text{ since $c \in \{0, 1\}$}) \\
& = \sum_{i=1}^{|Q|}\sum_{k=1}^{|\theta|}\frac{1} {{\theta_k}^2}P(C|Q_i,D=\delta,\theta_k)P(Q_i, d=\delta, \theta_k) - \mathbb{E}\left[R|d=\delta\right]^2 \\
& = \sum_{i=1}^{|Q|}\sum_{k=1}^{|\theta|}\frac{1}{{\theta_k}^2}P(R|Q_i,d=\delta)\theta_k P(Q_i, d=\delta, \theta_k) - \mathbb{E}\left[R|d=\delta\right]^2 \\
& = \sum_{i=1}^{|Q|}\sum_{k=1}^{|\theta|}\frac{1}{\theta_k}P(R|Q_i,d=\delta)P(Q_i, d=\delta, \theta_k) - \mathbb{E}\left[R|d=\delta\right]^2 \\
\end{aligned}
\end{equation}

Given that $E[R]^2 \le 1$ since $0 < R \le 1$, this variance can be large, commonly larger than the variance of $c$ which is bound at $\frac{1}{4}$. As such, $Var(\text{IPW-CTR})$ can be much larger than $Var(\text{CTR})$ since both are the variance of the aforementioned sequences divided by $n$.

This is clearest in the case of a document only shown in a fixed position $k$. For such a document:

\begin{equation}
\begin{aligned}
Var\left(\frac{c}{\theta_k}\right) = \frac{Var(c)}{{\theta_k}^2}
\end{aligned}
\end{equation}

\begin{equation}
\begin{aligned}
Var\left(\frac{c}{\theta_k}\right) \ge Var(c) && \text{since $\theta_k \le 1$}
\end{aligned}
\end{equation}

Probabilistic bounds on the IPS estimator are given in \cite{Schnabel2016}, which makes it clear how "we are paying for the unbiased-ness of IPS in terms of variability".

\subsection{Consequences for ranking}

If $\theta$ is only lower bound at 0, extreme position bias can lead to high variance estimates for IPW-CTR. This could be a problem for search engines that support long lists of documents for individual queries with severe position bias at low positions.

Furthermore, even for search engines with moderate bounds on search result length and moderate position bias, using IPW-CTR could lead to severe issues with ranking.

Consider a search engine with a constant influx of new documents, such that $n$ can be low. In the extreme case where $n = 1$ with a document shown at position $k$, $Var(\text{IPW-CTR}) = Var(\frac{c}{\theta_k}) = \frac{Var(c)}{{\theta_k}^2}$. IPW-CTR will equal $\frac{1}{\theta_k}$ with probability $P(R|q,d=\delta)\theta_k$, and 0 otherwise.

With enough (rare) documents, the possibility of an extreme IPW-CTR is not only plausible, but likely. If sorting on IPW-CTR or using it as a dominant feature in a machine learning algorithm, this implies we may often show a low relevance document at the top of a list. These rankings could be suboptimal and highly unintuitive for users. This problem could be persistent as long as there is a continual influx of new documents.

\subsection{Contrasting alternatives to IPW-CTR}

One option is to \textit{clip} the position bias weights, limiting how extreme their values can be and thus putting strict bounds on the variance of the IPS estimator \citep{Ionides2008, Bottou2012, Saito2019}. This can directly reduce variance, but with a direct trade-off of more bias \citep{Swaminathan2015}. Alternative features, such as SNIPS or IPW-COEC, reduce variance without having to choose a cut-off for snipping, and they naturally have continuous effects with respect to position bias.

SNIPS and IPW-COEC are superficially similar, in that both divide by a sum of terms derived from position bias factors. The terms in SNIPS are the reciprocal of the position bias factors. For documents that vary in the position bias they have been subject to, the SNIPS estimate may be dominated by the most extreme observations. This happens because individual terms of this sum can be unbounded, as $\lim_{\theta_k\to0} \frac{1}{\theta_k} = \infty$.

Since IPW-CTR and IPW-COEC are equivalent for documents with fixed positions, for such documents IPW-COEC and IPW-CTR share their high variance problem. This implies a risk of subpar ranking because of continual newer documents in the corpus.

For large $n$, IPW-COEC has the benefit of less sensitivity to individual clicks. Consider a document shown in multiple positions. The change in IPW-CTR with one more click at a position $k$ is $\frac{1}{\theta_k}$, while for IPW-COEC it is $\frac{1}{n \cdot \bar{\theta}}$. In the usual case of $n\cdot\bar{\theta} > \theta_k$, IPW-CTR sensitivity will be larger than IPW-COEC sensitivity.

\section{Experiments}

Comparing IPW-COEC and IPW-CTR:

\begin{itemize}
  \item Both are unbiased estimators of relevance
  \item They are equivalent for documents with fixed positions
  \item More generally, they are not equivalent
  \item They both have high variance for documents with low sample when position bias is significant
  \item IPW-COEC typically has less sensitivity to clicks at extreme positions
\end{itemize}

We designed artificial experiments to compare these two unbiased features in a realistic setting, and to contrast them against biased click-through rates. These experiments are illustrative, meant to demonstrate the dynamics described in this paper in a typical scenario, without claiming that the specific relationships will be equivalent in other empirical settings.

In our setting, these terms are model \textit{features}, rather than model \textit{estimators}. Practitioners can choose from a multitude of available estimators, while these historical rates can be important features provided to their models. While practitioners have to choose a single estimator, they can choose a multitude of features. Our work shows a fundamentally pluralistic view of these features, establishing how performance of each feature depends on the circumstances, such that practitioners may benefit from including a variety of such features in their models.

\subsection{Setup}

\subsubsection{The need for synthetic datasets}

We generate a novel synthetic dataset. Typical datasets in the learning to rank literature contain no document identifiers and have no discussion of document recurrence \citep{Chapelle10, Qin2010}. We want a setting rich enough to include realistic elements such as recurring documents and documents that are commonly shown together.

\subsubsection{Documents and user behavior}

We create a corpus of documents, where each document belongs to a cluster. The clusters are mutually exclusive and collectively exhaustive, and they vary substantially in the number of documents in each. Users search within a cluster, the search engine returns documents, and the user may then click on documents. Searches are distributed evenly and randomly across clusters.

User behavior follows the position-based propensity model, such that users only care about document relevance and position bias. We model position bias of the form $P(E|k) = k^x$, with a configuration of $x = 0.5$ except when varied to analyze sensitivity. This value was chosen to approximate the range of position bias curves demonstrated by known randomization experiments on large search engines \citep{Wang18, Agarwal19b, DemsynJones22}. 

Document relevance is chosen randomly and independently, with uniform probability between 0 and 1. The search engine does not know document relevance, but it does know a fairly powerful proxy that has high correlation with relevance which we use as an illustrative alternative to historical rate models. 

\subsubsection{Ranking and experimentation}

For each search the search engine picks 10 candidates randomly from the cluster and ranks them. This creates some variability in position for most documents, depending on the size of each cluster and whether the document has one of the largest or lowest ranking scores in its cluster. This mimics a realistic amount of variability per document, where in typical products a document may rank somewhat higher or lower due to availability, experiments, or other sources of variation, but the document typically does not rank far higher or lower than it usually would.

Our search engine first ranks by the relevance proxy for some number of searches. In subsequent searches, it chooses to rank using any of the aforementioned click rate estimates as a sole feature. We make no assumption on what type of model estimator is used, only assuming that these estimators would learn a monotonic relationship with respect to the single input feature.

For example, the search engine may rank using the relevance proxy for 10,000 searches (the "training" sample), then compute per-document click through rates (CTR's), with which it ranks for the following 100,000 searches (the "test" sample). We evaluate performance based on user click rates in the test sample, where more total clicks is better than fewer clicks. 

We repeat this for each feature, using the same test sample for each. We repeat this process 10 times, generating 10 test samples for any given configuration, estimating user click rates under ranking with each feature and averaging by feature.

To understand the dynamics with respect to sample size, we vary the size of the training sample, while keeping the test sample size fixed. We also vary the degree of position bias.

This design is meant to reasonably mimic typical search engines and user behavior. Relevance estimates are valuable as features to rank subsequent searches in order to maximize clicks. While a typical search engine may combine click-through rate features with many other features in a machine learning model, here we test their behavior without other confounding features.

More details on the experiment procedure are shared in Appendix \ref{appendix:simulation}.

\subsection{Results}

\subsection{IPW-CTR can approach the performance of true relevance and maximize clicks}

\begin{figure}[ht]
  \centering
  \includegraphics[width=350px]{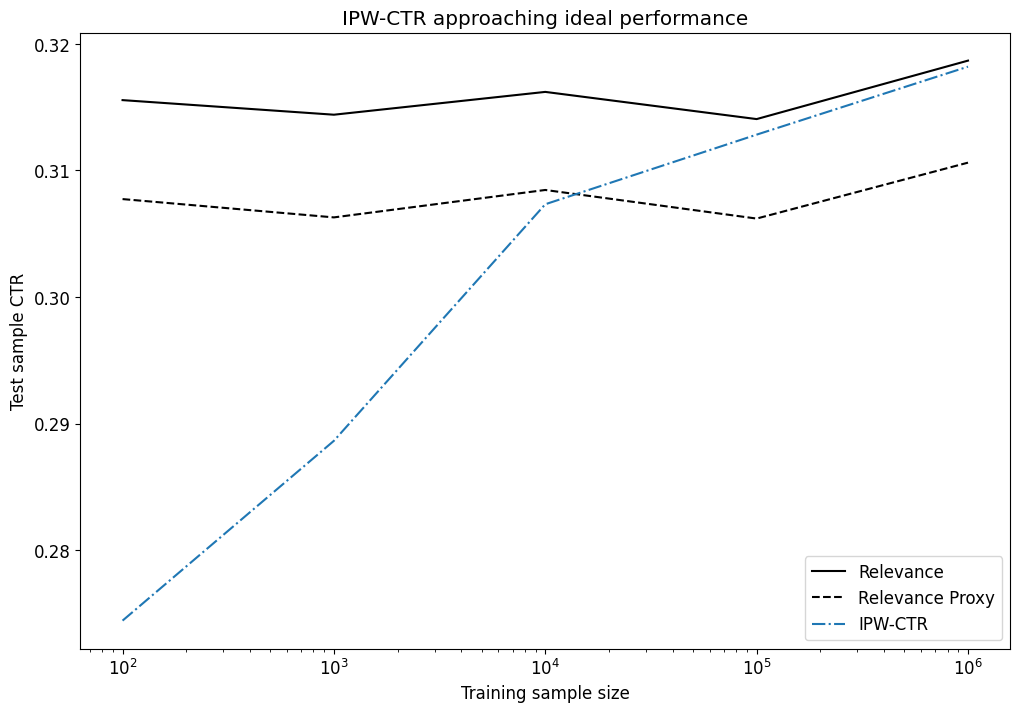}
  \caption{IPW-CTR is an unbiased estimator of relevance, and with enough sample can perform as well as relevance itself.}
  \label{fig:IPW-CTR}
\end{figure}

IPW-CTR is an unbiased estimator of relevance. While it cannot outperform a useful proxy until at least some data is accumulated, with enough sample per document it can near the performance of ranking by the oracle of true relevance, as demonstrated in Figure \ref{fig:IPW-CTR}. The relevance proxy does not improve as more sample is accumulated, since it is constant per document.

\subsection{CTR can outperform IPW-CTR}

\begin{figure}[ht]
  \centering
  \includegraphics[width=350px]{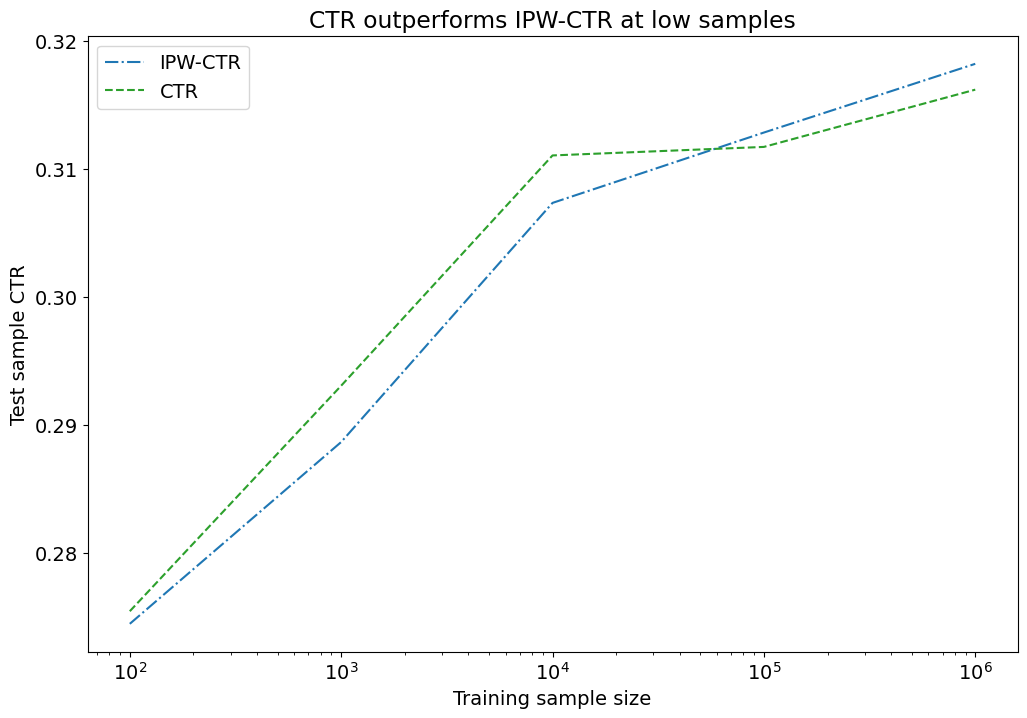}
  \caption{IPW-CTR's high variance hurts ranking performance at low sample sizes.}
  \label{fig:IPW-CTR_vs_CTR}
\end{figure}

Figure \ref{fig:IPW-CTR_vs_CTR} shows that click-through rate without any position adjustment at all can outperform IPW-CTR, despite CTR's bias. CTR has low variance, such that it can be a useful feature at lower sample sizes than IPW-CTR needs.

\begin{figure}[ht]
  \centering
  \includegraphics[width=450px]{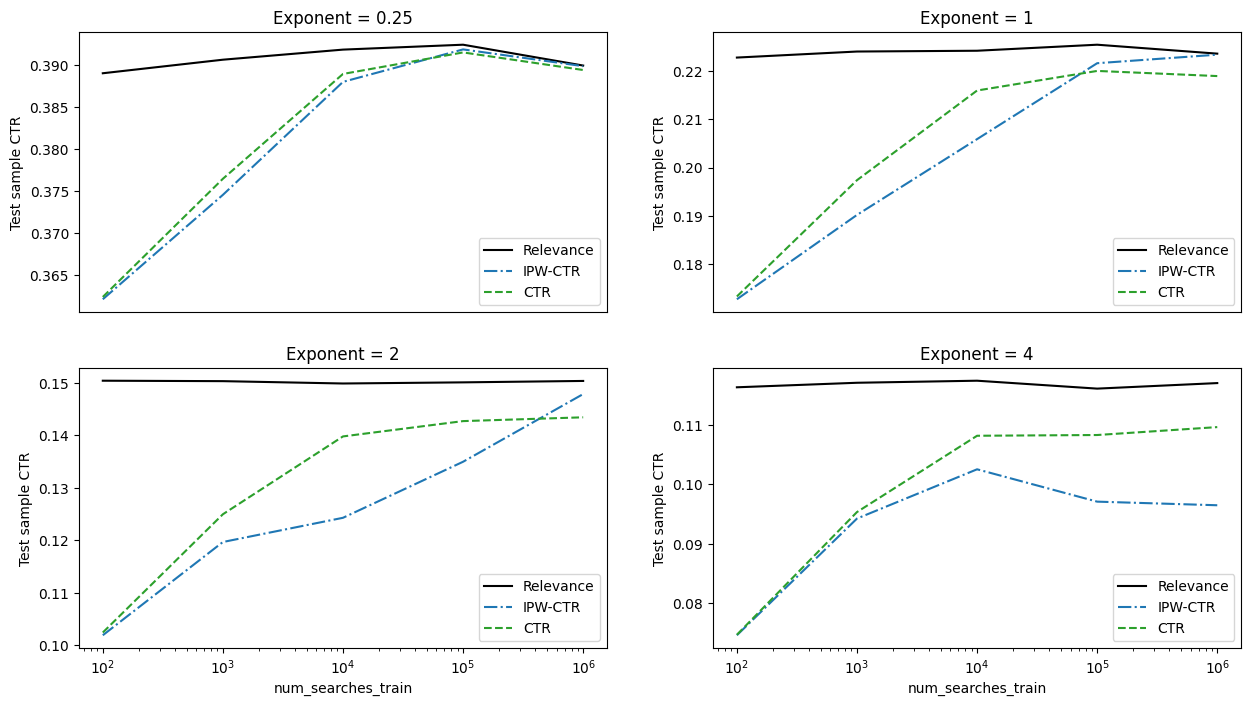}
  \caption{IPW-CTR's performance is variable at high degrees of position bias.}
  \label{fig:varying_exponents}
\end{figure}

Varying amounts of bias are shown in Figure \ref{fig:varying_exponents}.
At low degrees of position bias, even with large samples IPW-CTR may not outperform CTR, since the amount of bias in CTR will be limited. At high amounts of position bias, CTR can still sometimes outperform IPW-CTR, because of the high variance of IPW-CTR in such scenarios.

Two factors in this experiment design help CTR perform well, despite how it reinforces existing position biases:

\begin{itemize}
  \item We have meaningful random variation in rank
  \item Our initial ranking is based on a strong proxy of relevance
\end{itemize}

We made these choices to better mimic realistic scenarios. In empirical settings, contextual queries and iterations in ranking algorithms will typically lead to position variation for each document. We typically build models on logged data that used a somewhat reasonable ranker with at least some positive correlation with true relevance.

\subsection{IPW with biased (empirical) weights performs poorly}

\begin{figure}[ht]
  \centering
  \includegraphics[width=350px]{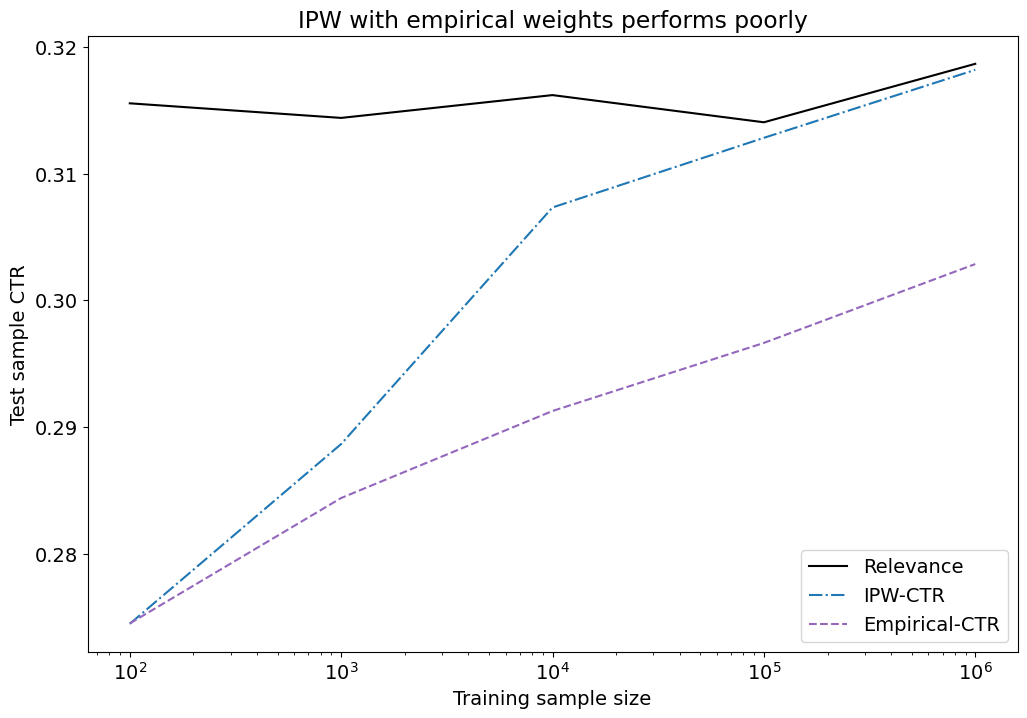}
  \caption{Empirical weights perform poorly.}
  \label{fig:empirical-ctr}
\end{figure}

Using average click-through rates by position as a proxy for position bias is extremely damaging. This feature does worse than both IPW-CTR and CTR, consistently across experiment specifications, with one shown in Figure \ref{fig:empirical-ctr}. Practitioners may be better off ignoring position bias than using a proxy that overestimates it.

\subsection{Choosing between COEC and CTR matters less than the choice of weights}

\begin{figure}[ht]
  \centering
  \includegraphics[width=350px]{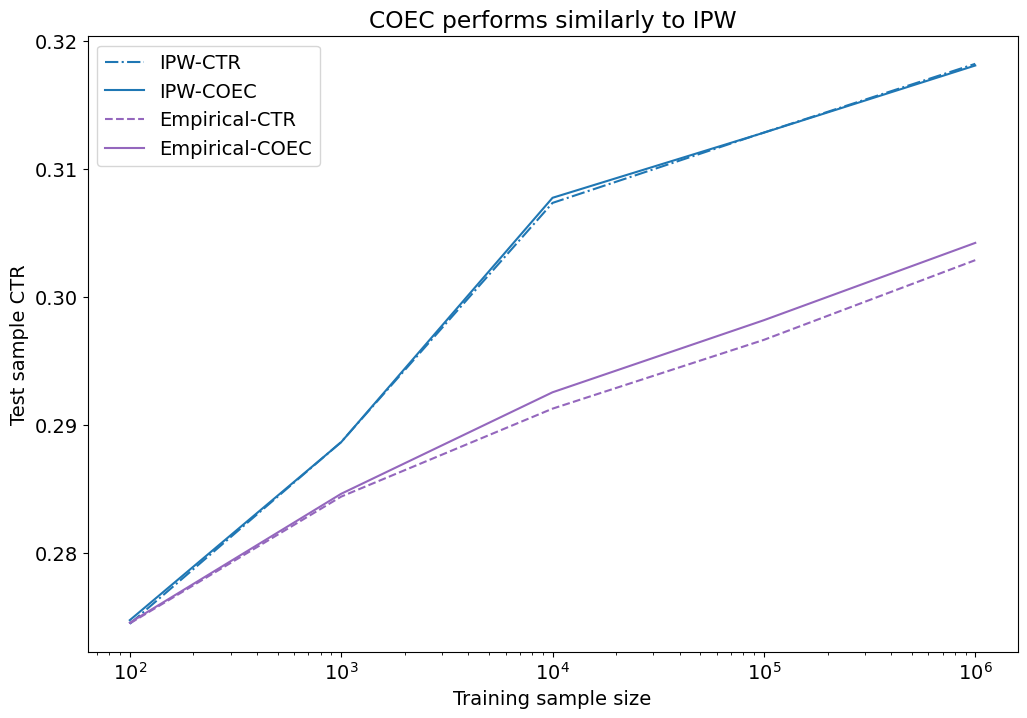}
  \caption{CTR or COEC weighting perform similarly.}
  \label{fig:IPW_vs_COEC}
\end{figure}

Figure \ref{fig:IPW_vs_COEC} is illustrative of a trend across various specifications, of IPW-CTR performing similarly to IPW-COEC, Empirical-CTR performing similarly to Empirical-COEC, and inverse propensity weights clearly out-performing empirical weights.

\begin{figure}[ht]
  \centering
  \includegraphics[width=300px]{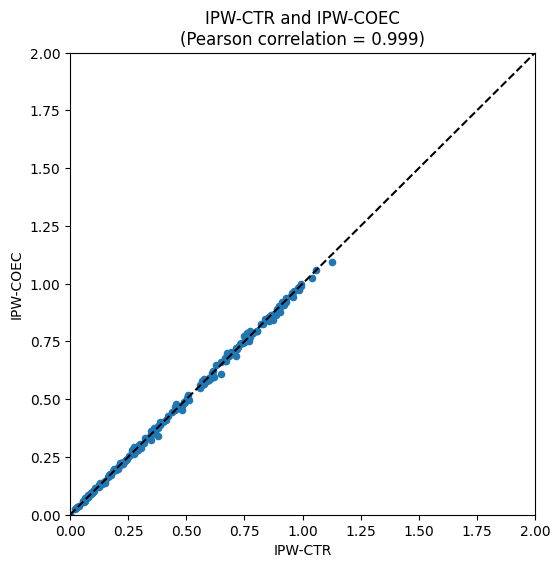}
  \caption{High correlation between IPW-CTR and IPW-COEC.}
  \label{fig:IPW_and_COEC_correlation}
\end{figure}

The similar performance of IPW-CTR and IPW-COEC is unsurprising because their values do not differ substantially at typical low position bias factors. Figure \ref{fig:IPW_and_COEC_correlation} demonstrates how the two features have similar absolute values and very high correlation. In our single-feature specification, ranking can only vary when relative ranking between observations differs between the two features.

\begin{figure}[ht]
  \centering
  \includegraphics[width=450px]{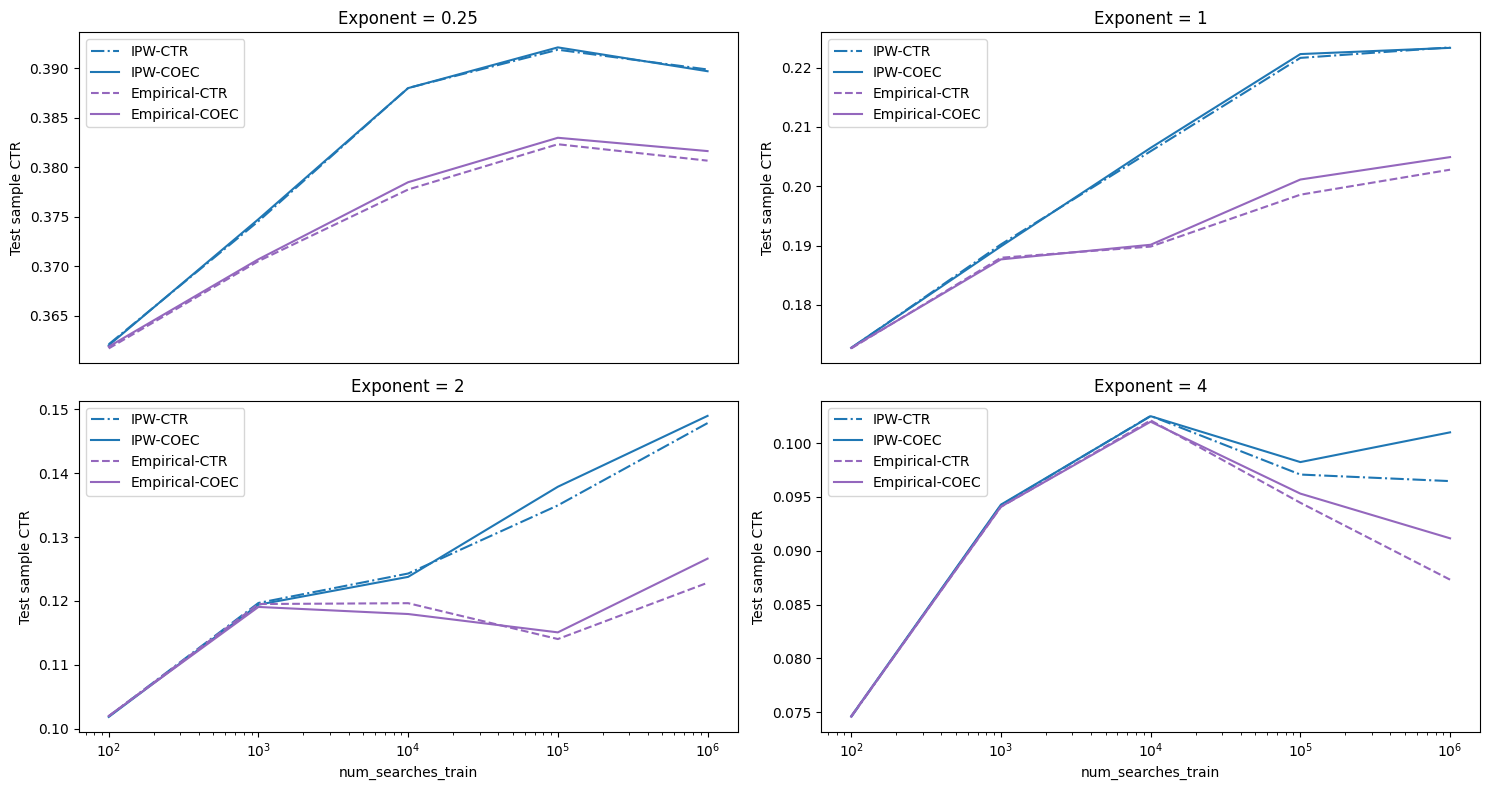}
  \caption{IPW-COEC's difference from IPW-CTR increases with position bias.}
  \label{fig:IPW_and_COEC_varying_coefficients}
\end{figure}

While performance is similar at typical position bias levels, COEC weighting seems to be an improvement at higher position bias levels where the variance difference between it and IPW-CTR can be larger, as shown in Figure \ref{fig:IPW_and_COEC_varying_coefficients}.

\section{Conclusion}

This paper illustrated important results through proofs and experiments. Extending the inverse propensity weighting (IPW, or IPS) method to documents creates an unbiased feature for document relevance (IPW-CTR). This feature can approximate relevance accurately, leading to near-optimal ranking. However, IPW-CTR has high variance, increasing with respect to the degree of position bias. Under some scenarios, biased click-through rate (CTR) can be a better feature than unbiased IPW-CTR.

Clicks over expected clicks (COEC) in its traditional empirically-weighted formulation can be a very poor feature for ranking. This is due to the overestimated position bias. Correct position bias weights for COEC (IPW-COEC) creates an unbiased estimator of relevance and often will have lower variance than IPW-CTR, such that it can be a better ranking feature. 

Practitioners may be well-advised to use IPW-COEC features, which have been only rarely discussed in the literature, and importantly to carefully estimate position bias factors. Furthermore, practitioners can choose a multitude of position-bias-adjusted features to include in their models, such as biased CTR, IPW-CTR, IPW-COEC, and SNIPS. Our paper is rare or unique in emphasizing this fundamentally pluralistic view of separating position bias adjustment in model estimators from position bias adjustment in features.

\section*{Acknowledgements}
We thank our colleagues for contributing to the initiatives that inspired this work, and for thoughtful insights in general. We particularly thank Denys Kopiychenko for detailed feedback on drafts of this paper.

\bibliographystyle{plainnat}
\bibliography{references}

\begin{thebibliography}{30}
\providecommand{\natexlab}[1]{#1}
\providecommand{\url}[1]{\texttt{#1}}
\expandafter\ifx\csname urlstyle\endcsname\relax
  \providecommand{\doi}[1]{doi: #1}\else
  \providecommand{\doi}{doi: \begingroup \urlstyle{rm}\Url}\fi

\bibitem[Agarwal et~al.(2018)Agarwal, Takatsu, Zaitsev, and
  Joachims]{Agarwal2018}
Aman Agarwal, Kenta Takatsu, Ivan Zaitsev, and Thorsten Joachims.
\newblock A general framework for counterfactual learning-to-rank.
\newblock \emph{Proceedings of the 42nd International ACM SIGIR Conference on
  Research and Development in Information Retrieval}, 2018.

\bibitem[Agarwal et~al.(2019{\natexlab{a}})Agarwal, Wang, Li, Bendersky, and
  Najork]{Agarwal19b}
Aman Agarwal, Xuanhui Wang, Cheng Li, Michael Bendersky, and Marc Najork.
\newblock Addressing trust bias for unbiased learning-to-rank.
\newblock \emph{The World Wide Web Conference}, 2019{\natexlab{a}}.

\bibitem[Agarwal et~al.(2019{\natexlab{b}})Agarwal, Zaitsev, Wang, Li, Najork,
  and Joachims]{Agarwal19}
Aman Agarwal, Ivan Zaitsev, Xuanhui Wang, Cheng Li, Marc Najork, and Thorsten
  Joachims.
\newblock Estimating position bias without intrusive interventions.
\newblock \emph{Proceedings of the Twelfth ACM International Conference on Web
  Search and Data Mining}, 2019{\natexlab{b}}.

\bibitem[Agichtein et~al.(2006)Agichtein, Brill, Dumais, and
  Ragno]{Agichtein2006}
Eugene Agichtein, Eric Brill, Susan~T. Dumais, and Robert~J. Ragno.
\newblock Learning user interaction models for predicting web search result
  preferences.
\newblock \emph{Proceedings of the 29th annual international ACM SIGIR
  conference on Research and development in information retrieval}, 2006.

\bibitem[Ai et~al.(2018)Ai, Bi, Luo, Guo, and Croft]{Ai2018}
Qingyao Ai, Keping Bi, Cheng Luo, J.~Guo, and W.~Bruce Croft.
\newblock Unbiased learning to rank with unbiased propensity estimation.
\newblock \emph{The 41st International ACM SIGIR Conference on Research \&
  Development in Information Retrieval}, 2018.

\bibitem[Aslanyan and Porwal(2019)]{Aslanyan19}
G.~Aslanyan and Utkarsh Porwal.
\newblock Position bias estimation for unbiased learning-to-rank in ecommerce
  search.
\newblock In \emph{SPIRE}, 2019.

\bibitem[Bottou et~al.(2012)Bottou, Peters, Candela, Charles, Chickering,
  Portugaly, Ray, Simard, and Snelson]{Bottou2012}
L{\'e}on Bottou, J.~Peters, Joaquin~Qui{\~n}onero Candela, Denis~Xavier
  Charles, David~Maxwell Chickering, Elon Portugaly, Dipankar Ray, Patrice~Y.
  Simard, and Edward Snelson.
\newblock Counterfactual reasoning and learning systems: the example of
  computational advertising.
\newblock \emph{ArXiv}, abs/1209.2355, 2012.

\bibitem[Chapelle and Chang(2010)]{Chapelle10}
Olivier Chapelle and Yi~Chang.
\newblock Yahoo! learning to rank challenge overview.
\newblock In \emph{Yahoo! Learning to Rank Challenge}, 2010.

\bibitem[Chapelle and Zhang(2009)]{Chapelle09}
Olivier Chapelle and Ya~Zhang.
\newblock A dynamic bayesian network click model for web search ranking.
\newblock In \emph{WWW '09}, 2009.

\bibitem[Cheng and Cant{\'u}-Paz(2010)]{Cheng2010}
Haibin Cheng and Erick Cant{\'u}-Paz.
\newblock Personalized click prediction in sponsored search.
\newblock In \emph{Web Search and Data Mining}, 2010.

\bibitem[Craswell et~al.(2008)Craswell, Zoeter, Taylor, and Ramsey]{Craswell08}
Nick Craswell, Onno Zoeter, Michael~J. Taylor, and Bill Ramsey.
\newblock An experimental comparison of click position-bias models.
\newblock In \emph{WSDM '08}, 2008.

\bibitem[Demsyn-Jones(2022)]{DemsynJones22}
Richard Demsyn-Jones.
\newblock Measurement and applications of position bias in a marketplace search
  engine, 2022.
\newblock URL \url{https://arxiv.org/abs/2206.11720}.

\bibitem[Dupret and Piwowarski(2008)]{Dupret08}
Georges Dupret and Benjamin Piwowarski.
\newblock A user browsing model to predict search engine click data from past
  observations.
\newblock In \emph{SIGIR '08}, 2008.

\bibitem[Guo et~al.(2019)Guo, Yu, Liu, Tang, and Zhang]{Guo19}
Huifeng Guo, Jinkai Yu, Qing Liu, Ruiming Tang, and Yuzhou Zhang.
\newblock Pal: a position-bias aware learning framework for ctr prediction in
  live recommender systems.
\newblock \emph{Proceedings of the 13th ACM Conference on Recommender Systems},
  2019.

\bibitem[Ionides(2008)]{Ionides2008}
Edward~L. Ionides.
\newblock Truncated importance sampling.
\newblock \emph{Journal of Computational and Graphical Statistics},
  17:\penalty0 295 -- 311, 2008.

\bibitem[Joachims et~al.(2007)Joachims, Granka, Pan, Hembrooke, Radlinski, and
  Gay]{Joachims07}
Thorsten Joachims, Laura~A. Granka, Bing Pan, Helene Hembrooke, Filip
  Radlinski, and Geri Gay.
\newblock Evaluating the accuracy of implicit feedback from clicks and query
  reformulations in web search.
\newblock \emph{ACM Trans. Inf. Syst.}, 25:\penalty0 7, 2007.

\bibitem[Joachims et~al.(2018)Joachims, Swaminathan, and Schnabel]{Joachims17}
Thorsten Joachims, Adith Swaminathan, and Tobias Schnabel.
\newblock Unbiased learning-to-rank with biased feedback.
\newblock In \emph{IJCAI}, 2018.

\bibitem[Liu et~al.(2017)Liu, Rogers, Shiau, Kislyuk, Ma, Zhong, Liu, and
  Jing]{Liu2017}
David~C. Liu, Stephanie Rogers, Raymond Shiau, Dmitry Kislyuk, Kevin C.~K. Ma,
  Zhigang Zhong, Jenny Liu, and Yushi Jing.
\newblock Related pins at pinterest: The evolution of a real-world recommender
  system.
\newblock \emph{Proceedings of the 26th International Conference on World Wide
  Web Companion}, 2017.

\bibitem[Ovaisi et~al.(2020)Ovaisi, Ahsan, Zhang, Vasilaky, and
  Zheleva]{Ovaisi20}
Zohreh Ovaisi, Ragib Ahsan, Yifan Zhang, Kathryn~N. Vasilaky, and Elena
  Zheleva.
\newblock Correcting for selection bias in learning-to-rank systems.
\newblock \emph{Proceedings of The Web Conference 2020}, 2020.

\bibitem[Qin and Liu(2010)]{Qin2010}
Tao Qin and Tie{-}Yan Liu.
\newblock Microsoft learning to rank datasets.
\newblock \emph{Microsoft Research}, 2010.
\newblock URL \url{https://www.microsoft.com/en-us/research/project/mslr/}.

\bibitem[Saito et~al.(2019)Saito, Yaginuma, Nishino, Sakata, and
  Nakata]{Saito2019}
Yuta Saito, Suguru Yaginuma, Yuta Nishino, Hayato Sakata, and Kazuhide Nakata.
\newblock Unbiased recommender learning from missing-not-at-random implicit
  feedback.
\newblock \emph{Proceedings of the 13th International Conference on Web Search
  and Data Mining}, 2019.

\bibitem[Schnabel et~al.(2016)Schnabel, Swaminathan, Singh, Chandak, and
  Joachims]{Schnabel2016}
Tobias Schnabel, Adith Swaminathan, Ashudeep Singh, Navin Chandak, and Thorsten
  Joachims.
\newblock Recommendations as treatments: Debiasing learning and evaluation.
\newblock \emph{ArXiv}, abs/1602.05352, 2016.

\bibitem[Schroedl et~al.(2010)Schroedl, Kesari, and Neumeyer]{Schroedl2010}
Stefan Schroedl, Anand Kesari, and Leonardo Neumeyer.
\newblock Personalized ad placement in web search.
\newblock \emph{ADKDD}, 2010.

\bibitem[Swaminathan and Joachims(2015)]{Swaminathan2015}
Adith Swaminathan and Thorsten Joachims.
\newblock The self-normalized estimator for counterfactual learning.
\newblock In \emph{NIPS}, 2015.

\bibitem[Teo et~al.(2016)Teo, Nassif, Hill, Srinivasan, Goodman, Mohan, and
  Vishwanathan]{Teo2016}
Choon~Hui Teo, Houssam Nassif, Daniel~N. Hill, Sriram Srinivasan, Mitchell
  Goodman, Vijai Mohan, and S.~V.~N. Vishwanathan.
\newblock Adaptive, personalized diversity for visual discovery.
\newblock \emph{Proceedings of the 10th ACM Conference on Recommender Systems},
  2016.

\bibitem[Vardasbi et~al.(2020)Vardasbi, Oosterhuis, and de~Rijke]{Vardasbi20}
Ali Vardasbi, Harrie Oosterhuis, and M.~de~Rijke.
\newblock When inverse propensity scoring does not work: Affine corrections for
  unbiased learning to rank.
\newblock \emph{Proceedings of the 29th ACM International Conference on
  Information \& Knowledge Management}, 2020.

\bibitem[Wang et~al.(2018)Wang, Golbandi, Bendersky, Metzler, and
  Najork]{Wang18}
Xuanhui Wang, Nadav Golbandi, Michael Bendersky, Donald Metzler, and Marc
  Najork.
\newblock Position bias estimation for unbiased learning to rank in personal
  search.
\newblock \emph{Proceedings of the Eleventh ACM International Conference on Web
  Search and Data Mining}, 2018.

\bibitem[Yan et~al.(2022)Yan, Qin, Zhuang, Wang, Bendersky, and Najork]{Yan22}
Le~Yan, Zhen Qin, Honglei Zhuang, Xuanhui Wang, Mike Bendersky, and Marc
  Najork.
\newblock Revisiting two tower models for unbiased learning to rank.
\newblock \emph{Proceedings of the 45th International ACM SIGIR Conference on
  Research and Development in Information Retrieval}, 2022.

\bibitem[Zhang and Jones(2007)]{Zhang2007}
Wei~Vivian Zhang and R.~Jones.
\newblock Comparing click logs and editorial labels for training query
  rewriting.
\newblock In \emph{WWW '07}, 2007.

\bibitem[Zhang et~al.(2022)Zhang, Yan, Qin, Zhuang, Shen, Wang, Bendersky, and
  Najork]{Zhang22}
Yunan Zhang, Le~Yan, Zhen Qin, Honglei Zhuang, Jiaming Shen, Xuanhui Wang,
  Michael Bendersky, and Marc Najork.
\newblock Towards disentangling relevance and bias in unbiased learning to
  rank.
\newblock \emph{ArXiv}, abs/2212.13937, 2022.

\end{thebibliography}

\appendix

\section{Simulation procedure}
\label{appendix:simulation}

The general procedure for this paper is described in Algorithm \ref{alg:simulate}. This is a simplification of the actual code, omitting detail and vectorization.

\begin{algorithm}
\caption{Simulation algorithm (iterative)}\label{alg:simulate}
\begin{algorithmic}
\State $parameter\_set \gets \text{size and exponent conditions}$
\State $feature\_set \gets \text{candidate CTR features}$
\State $proxy\_feature \gets \text{relevance proxy}$
\ForAll{$parameters$ $\in$ $parameter\_set$}
    \For{$1$ to $10$}
        \State $clusters \gets \text{generate\_clusters($parameters$)}$
        \State $searches\_train, searches\_test \gets \text{generate\_searches($parameters$, $clusters$)}$
        \ForAll{$search$ $\in$ $searches\_train$} \text{// In place modifications}
            \State $search \gets \text{rank\_by($search$, $proxy\_feature$)}$
            \State $search \gets \text{simulate\_user\_behavior($search$)}$
        \EndFor
        \State $features \gets \text{calculate\_features($searches\_train$, $feature\_set$)}$
        \State $searches\_test \gets \text{apply\_features($searches\_test$, $features$)}$
        \ForAll{$feature$ $\in$ $feature\_set$}
            \ForAll{$search$ $\in$ $searches\_test$} \text{// In place modifications}
                \State $search \gets \text{rank\_by($search$, $feature$)}$
                \State $search \gets \text{simulate\_user\_behavior($search$)}$
            \EndFor
            \State \text{calculate\_and\_log\_performance($searches\_test$)}
        \EndFor
    \EndFor
    \State \text{calculate\_and\_log\_average\_performance()}
\EndFor
\State \text{summarize\_all\_experiments()}
\end{algorithmic}
\end{algorithm}

For this paper, we define a rich setting quite different from most of the literature, allowing for substantial variation in volume and ranked position per document while also having recurring pairs of documents shown together with fixed relative ordering in the training sample.

We have a concept of \textit{clusters}. A cluster is a set of documents that could be shown together in searches. Clusters are non-overlapping. Imagine a search product with mutually exclusive categories in which users can search and filter. Clusters vary in size, yet the search volume per cluster is unrelated to cluster size. Clusters of size 0 are not permitted or searchable. Regardless of cluster size, the search engine returns at most a fixed maximum of documents per search.

\begin{figure}[ht]
  \centering
  \includegraphics[width=350px]{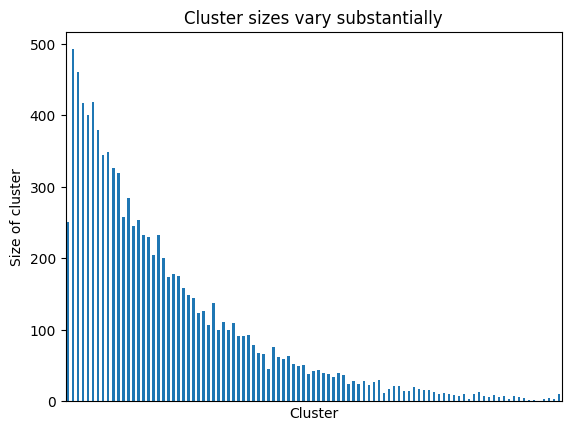}
  \caption{Size of cluster following an exponential distribution.}
  \label{fig:cluster_sizes}
\end{figure}

Let's consider an example of 100 clusters averaging 100 documents each, a maximum of 10 documents returned per search, and 10,000 searches. As shown in Figure \ref{fig:cluster_sizes}, we have some large clusters (approaching 500 documents), some small clusters (of fewer than 10 documents), and a range of cluster sizes in between.

\begin{figure}[ht]
  \centering
  \includegraphics[width=350px]{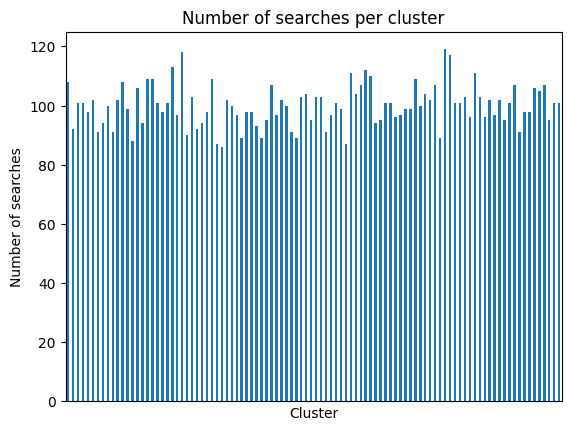}
  \caption{Each cluster randomly selected for approximately 100 searches.}
  \label{fig:searches_per_cluster}
\end{figure}

With 10,000 searches that choose independently and uniformly from clusters, our search volume per cluster will approximate 10,000 / 100 = 100, as demonstrated in Figure \ref{fig:searches_per_cluster}.

\begin{figure}[ht]
  \centering
  \includegraphics[width=350px]{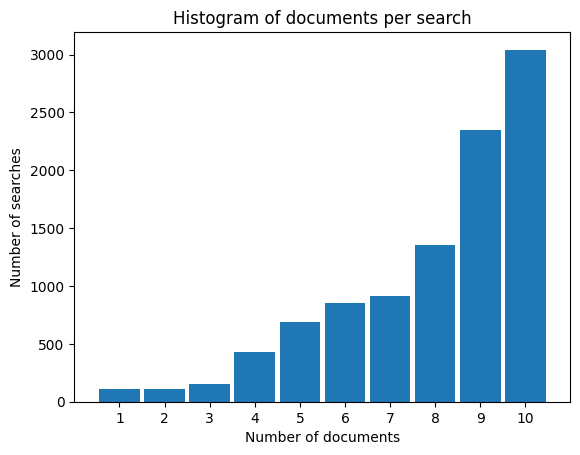}
  \caption{Each search has up to 10 documents, depending on cluster size.}
  \label{fig:documents_per_search}
\end{figure}

Since no more than 10 documents are returned per search, and most clusters have more than 10 documents, we will most often have exactly 10 documents returned in searches. However, smaller returned lists will be plentiful in the data, including a small but non-zero amount of lists with only one or two documents. Figure \ref{fig:documents_per_search} shows the commonality of search result sizes.

Searches in large clusters will return 10 documents per search, and each of those clusters will be searched approximately 100 times, representing approximately 1,000 records (search-document pairs). Smaller clusters will be represented in fewer total records. Given the amount of simulations and data generated, we vectorize the procedure in a way that can create duplicate records, which we then drop. While this results in a different distribution than an iterative process would, it is unclear which distribution is more or less realistic compared to any specific real-world setting. Either distribution will have suitable variation and richness to demonstrate the properties of this paper.

\begin{figure}[ht]
  \centering
  \includegraphics[width=350px]{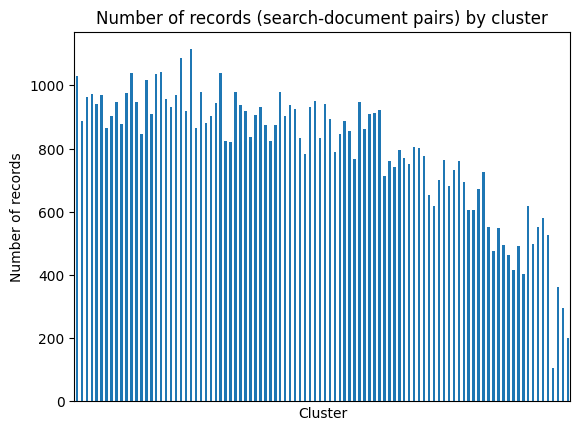}
  \caption{Larger clusters will account for more total records than small clusters.}
  \label{fig:records_per_cluster}
\end{figure}

Figure \ref{fig:records_per_cluster} shows the resulting number of records per cluster in this example.

This paper demonstrates results over a variety of data set sizes and behavioral parameters, with clusters, searches, and user behavior on those searches generated randomly anew for each parameter set. For every generated dataset, each feature is evaluated, rather than randomizing the data separately for each feature evaluation.

\end{document}